# Magnetic filaments: formation, stability, and feedback

*Evgeny A. Kuznetsov[1,2,3,4], Evgeny A. Mikhailov[1,2,5]*
[1]*P.N.Lebedev Physical Institute, 119991 Moscow, Russia*
[2]*Skolkovo Institute of Science and Technology, 121205 Moscow, Russia*
[3]*L.D.Landau Institute for Theoretical Physics, 142432 Chernogolovka, Russia*
[4]*Space Research Institute, 117997 Moscow, Russia*
[5]*M.V.Lomonosov Moscow State University, 119991 Moscow, Russia*
e-mail: kuznetso@itp.ac.ru, ea.mikhajlov@physics.msu.ru

As well known, magnetic fields in space are distributed very inhomogeneously. Some-times field distributions have forms of filaments with high magnetic field values. As many ob-servations show, such a filamentation takes place in convective cells in the Sun and other astro-physical objects. This effect is associated with the frozenness of the magnetic field into a medium with high conductivity that leads to compression of magnetic field lines and forming magnetic filaments. We show analytically, based on the general analysis, that the magnetic field intensifies in the regions of downward flows in both two-dimensional and three-dimensional convective cells. These regions of the hyperbolic type for magnetic fields play a role of a specific attractor. This analysis was confirmed by numerical simulations for 2D convective cells of the roll-type. Without dissipation the magnetic field grows exponentially in time and does not depend on the aspect ratio between horizontal and vertical scale of the cell. An increase due to compression in the magnetic field in the high conductive plasma is saturated due to the natural limitation associated with dissipative effects when the maximum magnitude of the magnetic field is of the order of the root of the magnetic Reynolds number Rem. For the solar convective zone the mean kinetic energy density exceeds mean magnetic energy density at least for two orders of magnitude that allows one to use the kinematic approximation for the MHD induction equation. In this paper based on the stability analysis we explain why downward flows influence magnetic filaments from making them more flat with orientation along interfaces between convective cells.

## 1. Introduction

The phenomenon of collapse plays a significant role in terms of understanding how turbulence, convection and other similar phenomena operate in fluids. The collapse is understood as a process of formation of singularities in a finite time for smooth initial conditions. Such processes have been widely studied for quite a long time. According to classical concepts of the Kolmogorov - Obukhov theory [1, 2] in the case of a low viscosity limit, the vorticity fluctuations in the inertial interval with a scale $\lambda$ behave proportionally to $\lambda^{-2/3}$. This means that in the limit of small $\lambda$ we will have infinite amplitudes of fluctuations, which may indicate that classical turbulence is closely related to the occurrence of collapse. At the same time, when highly accurate numerical modeling of such problems became possible, it turned out that in fact collapse was not observed in such cases [3] (see also the review paper [4] devoted to this subject). Nevertheless, the tendency for vorticity enhancement remains but without blow-up behavior. At the same time, for two-dimensional hydrodynamics in the ideal case, solutions associated with collapse are forbidden [5, 6, 7]. In the 2D case this, however, does not exclude the existence of infinite exponential growth [8]. In the 3D hydrodynamics, the generation of the pancake-type structures was numerically observed, when the pancake thickness decreases with time according to an exponential law due to compressible character of the vorticity field [9]. It is important to note that the appearance of such structures in the three-dimensional case is associated with the vorticity frozenness into a fluid [10]. It should be noted that in two-dimensional ideal flows, the vorticity rotor (called divorticity) is also a frozen vector field [8]. This means that not only vorticity, but also any other fields frozen into a medium should be compressible and all arguments about collapse given above are applicable to them. Compressible feature of such fields is a sequence of their frozenness.

Another classical example of frozen fields is a magnetic field in the ideal magnetohydrodynamics (MHD) [11]. In this case, we may expect that the magnetic field should evolve by the same laws as the vorticity in ideal fluids and consequently compress into localized magnetic structures [12]. But, unlike fluids, the MHD operates with two vector fields, namely, velocity and magnetic field. If kinetic energy density sufficiently exceeds magnetic energy density we can consider evolution of the magnetic field in a given velocity distribution and ignore the influence of the growing magnetic field on velocity flows. Such a situation is realized in the convective zone of the Sun where the ratio between mean kinetic energy density and magnetic pressure at least consists of two orders of magnitude. Indeed, in this case, the size of magnetic distributions has a tendency to decrease exponentially in time with exponential increase of the magnetic field values. Amplification of magnetic field is a direct consequence of its frozenness.

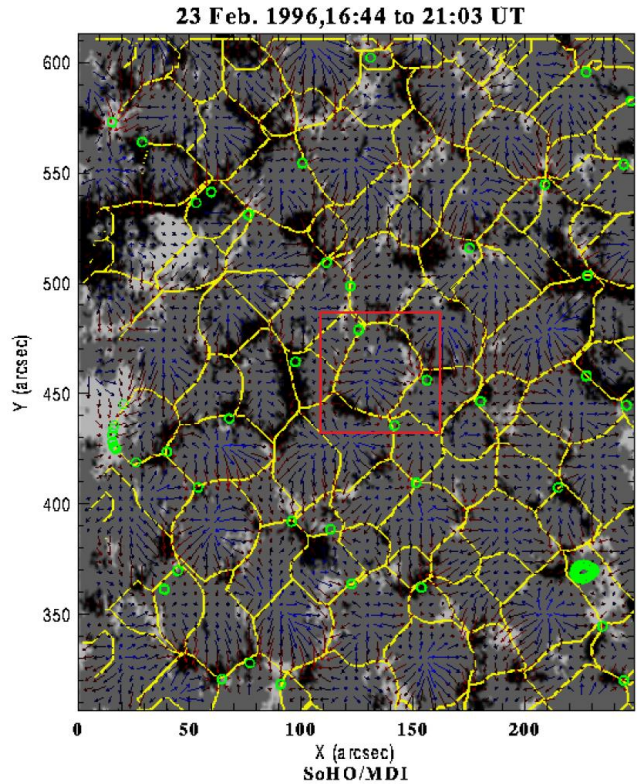

Figure 1. SOHO magnetogram overlaid with lines of convergence of the horizontal flow and with green dots showing the convergence points. The measured flow is shown as colored arrows, red for inferred downflow and blue for inferred upflow. The field is shown light grey for positive fields and dark for negative fields. Only field above the background noise is shown. Red rectangle shows the fragment which is demonstrated on fig.2. Courtesy of SOHO/MDI consortium. SOHO is a project of international cooperation between ESA and NASA [26].

From the astrophysical point of view, the process of appearance of localized strong magnetic fields has a great interest. Thus, for the first time Parker drew attention to the fact that the magnetic field in the convective zone of the Sun is noticeably filamented [13]. He studied the magnetic field evolution in the case of a two-dimensional velocity field corresponding to convective rolls and showed exponential in time generation of magnetic filaments. Subsequently, these ideas were developed in many other works on this topic (see, [14-25] and references therein). According to the data of Solar and Heliospherical Observatory (SOHO), in the convective zone of the Sun filaments in their majority are concentrated near interfaces between convective cells and oriented correspondingly to downward flows [26]. The SOHO observations also show that filaments are almost absent at centers of cells (Fig.1, Fig.2).

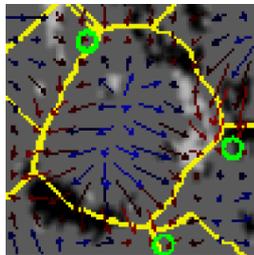

Figure 2. Single cell on the Sun (enlarged fragment of the magnetogram from Fig. 1). Courtesy of SOHO/MDI consortium. SOHO is a project of international cooperation between ESA and NASA [26].

An exponential increase in the magnetic field in the regions of downward flows follows from the topological arguments based on the Okubo–Weiss criteria [27, 28]. These criteria divide flow area by hyperbolic and elliptic parts. Downward flows belong to hyperbolic regions. Due to frozenness of the magnetic field into plasma these regions play a role of specific attractors. In elliptic regions magnetic fields demonstrate a formation of spiral structures. Accounting of finite conductivity in the hyperbolic regions leads to saturation of magnetic field growth up to values proportional to the root square of the magnetic Reynolds number $Re_m$ [14].

It is worth noting that both observations and estimations show that, in the convective zone of the Sun, mean kinetic energy is sufficiently large in comparison with magnetic energy, so we can use for description of magnetic field dynamics by the MHD equations in the kinematic approximation.

Note that processes associated with the influence of convective flows are no less important in other astrophysical objects, such as galaxies, accretion disks, etc. In the simplest case, they can be taken into account using averaged models with account of the helicity fluxes of the magnetic field. At the same time, it is necessary to more accurately consider the corresponding effects within the framework of numerical modeling of flows in such objects.

Previously, we studied the basic features of magnetic field amplification in convective cells [29] for the example of simple flow of the roll type within a two-dimensional model. It was shown both numerically and analytically that, indeed, in the vicinity of hyperbolic points of the flow, the magnetic field increases exponentially in time. In the ideal case ($\text{Re}_m \to \infty$) the field growth is not limited. For finite magnetic viscosity this process saturates that gives the resulting magnetic field values of the order of $\text{Re}_m^{1/2}$. It is also worth noting that in [29] we analyzed cells of a square shape, i.e., for equal horizontal and vertical cell sizes. In reality, such an assumption is unlikely to be realized. It should be thought that the vertical dimension is greater than the horizontal dimension. The fact is that the convective transfer of heat into the Sun extends, according to all estimates, to distances greater than the horizontal dimensions of the convective cells. The main thing is that we do not know at what depths the transition from turbulent convection to the observed laminar convection occurs. In this regard, it is necessary to understand how the process of formation of magnetic filaments is affected by an increase in the vertical size of the cells.

At the same time, the question arises as to whether the process of the filamentation can be limited by other mechanisms. It is evident that the kinematic approximation in [29] can not be applied to study the feedback influence of the growing magnetic field on the flow itself. Note that such approach is very popular and used often in dynamo theory [30, 31, 32]. In the case of dynamo theory, the simplest methods of nonlinear saturation are often used, which requires a separate study of the stability of stationary solutions. In our case, intensive magnetic fields lead to the emergence of a large Lorentz force in a vicinity of magnetic filaments. It will change the nature of the flow of the medium, which, according to Lentz's rule, can lead to a weakening of growth and probably its complete stop. Nevertheless, in the vicinity of a stationary hyperbolic point, the main direction of the Lorentz force will be parallel to the top convection surface with a high degree of accuracy. This means that the influence of the magnetic field on the flow may be strong enough in comparison with simple estimations. However, all this needs additional verification which is one of the main aims of this work.

In this work, we will show how the problem of the formation of magnetic filaments in the convective zone of the Sun can be qualitatively studied based only on an analysis of the dynamics of the free surface of convective cells, i.e. reducing the dimension of the problem. This makes it possible to explain the formation of magnetic filaments in the vicinity of the interfaces between convective cells, i.e. in areas of downward flows, which act as specific attractors for the magnetic field. The growth of the magnetic field in the filaments occurs due to the freezing of the field. It is important that this result does not depend on the specific structure of the convective cell. A numerical experiment confirms this analytical concept.

## 2. Parameters of solar convective zone

First of all, we present main parameters of the solar convective zone. According to observations, horizontal size $L$ for convective cells consists of 500–1000 km. For solar convective cells in this paper we will keep an assumption that ratio between such scales is of the same order of magnitude (but not infinitely large) as for laboratory convection [29] (see also [30]). For the solar convection zone kinetic energy density much larger magnetic energy density, $\frac{\rho \upsilon^2}{2} \gg \frac{B^2}{8\pi}$ (their ratio is about $10^3 - 10^4$) where $\rho$ is a mass density, $B$ is the magnetic field value and $\vec{\upsilon}$ a flow velocity being locally incompressible, $\text{div}\vec{\upsilon} = 0$. Therefore a fluid velocity in this case can be considered as a given vector field. This limit is known as a kinematic approximation which is widely used in the dynamo problem. Note also that near the boundary of the convective zone with photosphere (the beginning of the Sun atmosphere) mass density $\rho$ there is about $10^{-5}$ g/cm$^3$. The characteristic velocity $U$ for cells is about 500 m/s. There mean magnetic field $B_0$ is about a few gauss (for estimate we will take $B_0 = 10$ G).

Thus, dynamics of the magnetic field in the convective zone can be analyzed within one equation of the MHD system, i.e. for the magnetic field this is the induction equation:

$$\frac{\partial \vec{B}}{\partial t} = \vec{\nabla} \times (\vec{\upsilon} \times \vec{B}) + \frac{1}{\text{Re}_m} \Delta \vec{B} \qquad (1)$$

with a given (independent on magnetic field) velocity distribution $\vec{\upsilon}$ satisfying incompressibility condition $\text{div}\vec{\upsilon} = 0$.

In this equation $\text{Re}_m = \frac{UL}{\eta}$ is the magnetic Reynolds number and $\eta = \frac{c^2}{4\pi\sigma}$ is magnetic viscosity, where $c$ is the speed of light, $\sigma$ conductivity. According to all known data (see, e.g. [19, 20] and references there) in convective cells $\text{Re}_m$ is of order $10^6$, that allows one to neglect dissipation in equation (1). Thus, in this approximation the magnetic field $\vec{B}$ turns out to be frozen-in-fluid vector field. The frozenness means that magnetic field lines move due to the velocity component normal to the magnetic field direction. As follows from (1) in this case the parallel velocity component does not influence on a motion of magnetic field lines.

### 3. Generation of magnetic filaments by two-dimensional flows

In this section firstly a convective flow will be considered to be two-dimensional (lies in the $xy$ plane), stationary and periodic along horizontal coordinate $x$. Thus, cells are supposed to be of the roll type where flows in two neighboring cells have opposite rotations. The vertical coordinate of the cell lattice changes in the interval: $0 \leq y \leq h$; $y = 0$ corresponds to the upper boundary of the convective zone and $h$ is its depth. Such geometry of the convective flow means that normal velocity component $\upsilon_y = 0$ at $y = 0$, namely, we neglect by any perturbations of the top convective plane. Hence for $y$-component of the magnetic field at $y = 0$, in accordance with Eq. (1) and incompressibility condition $\text{div}\vec{\upsilon} = 0$, one can get the following equation

$$\frac{\partial B_y}{\partial t} + \upsilon_x \frac{\partial B_y}{\partial x} = -B_y \frac{\partial \upsilon_x}{\partial x} + \text{Re}_m^{-1} \nabla^2 B_y \qquad (2)$$

Note that this equation becomes autonomous at zeroth magnetic viscosity, namely,

$$\frac{\partial B_y}{\partial t} + \upsilon_x \frac{\partial B_y}{\partial x} = -B_y \frac{\partial \upsilon_x}{\partial x}. \qquad (3)$$

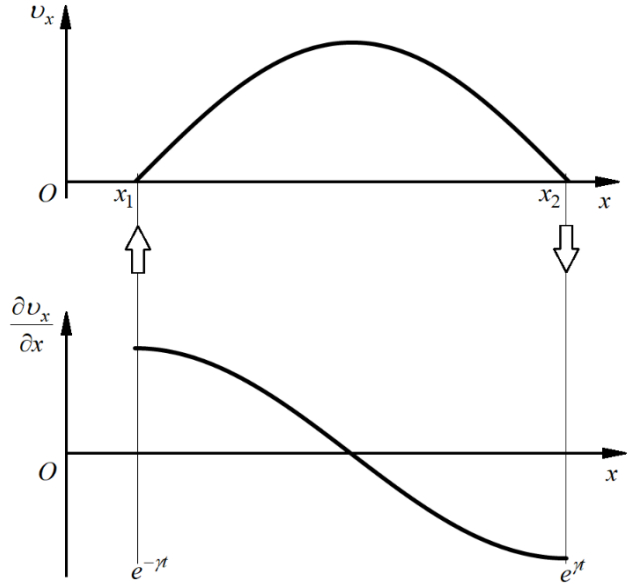

Figure 3. Scheme of the generation of the magnetic field. Upper plot shows the profile of the velocity component parallel to the surface $y = 0$, lower one – its spatial derivative. Arrows indicate centers of upperward and downward flows, respectively.

Consider the latter case in more details. In convection cell the parallel velocity component $\upsilon_x$ at $y = 0$ is equal to zero in two points where convective flow makes turns on $90^O$. These points correspond to centers for upward and downward flows. Hence it follows that derivatives of $\upsilon_x$ in these points have different signs (fig. 3). At the center of upward flow this derivative will be positive because this point ($x_1$) corresponds to flow sourse on the plane $y = 0$, but the second point ($x_2$) surves for sink and derivative there of $\upsilon_x$ will be negative. Such behavior of the parallel velocity on the plane $y = 0$ immediately gives that at the point with positive derivative $B_y$ will decrease exponentially in time: $\propto \exp\left\{-\left.\frac{\partial \upsilon_x}{\partial x}\right|_{x=x_1} t\right\}$, and increase exponentially in the sink point $x_2$: $\propto \exp\left\{\left.\frac{\partial \upsilon_x}{\partial x}\right|_{x=x_2} t\right\}$. If the parallel velocity symmetric (see Fig. 3) then it is evident that decrement $\gamma(x_1)$ and increment $\gamma(x_2)$ will have the same values but opposite signs. Note that between these two points as follows from equation (3) besides amplification the normal component of $\vec{B}$ undergoes advection towards point $x_2$. Note that this motion is a consequence of the frozenness property of magnetic field when the only $\vec{B}$-component perpendicular to the convective flow undergoes advection.

Consider now what happens with another magnetic component. As we saw above behavior of the $B_y$ component at $y = 0$ is based on both stationary character of the convective flow and assumption of the top convection boundary being planary. The same issue can be applied to the downward flow (plane $x = x_2$). Then evidentely that the equation of motion for $B_x$ there will be analogous to (2):

$$\frac{\partial B_x}{\partial t} + \upsilon_y \frac{\partial B_x}{\partial y} = -B_x \frac{\partial \upsilon_y}{\partial y}. \tag{4}$$

Here, at $y = 0$ the $x$-component of $\vec{\upsilon}$ will be equal to zero but its derivative at $x = x_2$ due to incompressibility condition will be positive:

$$\frac{\partial \upsilon_y}{\partial y} = -\frac{\partial \upsilon_x}{\partial x} > 0. \tag{5}$$

Hence, we have exponential in time decrease of $B_x$ with maximal decrement

$$\gamma_{dis} = -\left.\frac{\partial \upsilon_x}{\partial x}\right|_{x=x_2, y=0}. \tag{6}$$

Thus, in the corner ($y = 0$, $x = x_2$) the magnetic field grows exponentially in time where the magnetic field direction is parallel to the downward flow and respectively $|B_x| \ll |B_y|$. Note, that the point ($y = 0$, $x = x_2$) for this flow represents a hyperbolic point, respectively, the region around this point belongs to the hyperbolic one, in accordance with the Okubo-Weiss criterion (see below). Hence using the magnetic flux conservation it is possible to estimate width $\delta$ of the forming magnetic filament:

$$\langle B|_{t=0} \rangle L \propto B_{\max} \delta. \tag{7}$$

Here $\langle B|_{t=0} \rangle$ is the mean magnetic field and $L$ is the longitudinal size of the convective cell. This estimation gives that the filament width compresses in time exponentially with decrement $\gamma_{dis}$.

It is worth noting that the above analysis is general. It does not use a concrete structure of convective cell. The only assumption made is that a top convection surface is planar and a flow is stationary.

**4. Numerical simulations**

In numerics presented below we show that these general conjectures are completly valid in the partial case of a convective flow of the roll type. Consider the flow of this type written in terms of stream function $\psi$:

$$\upsilon_x = -\frac{\partial \psi}{\partial y}; \quad \upsilon_y = \frac{\partial \psi}{\partial x}, \tag{8}$$

with the choice $\psi$ in the form:

$$\psi(x, y) = V \sin k_1 x \sin k_2 y. \tag{9}$$

(Note that here we use for stream function different sign in comparison with the common definition that was more convinient for us in numerics.) Using dimensionless variables it is convenient to take values $V = \alpha$, $k_2 = 1/\alpha$ and $k_1 = 1$ where $\alpha$ is the aspect ratio, it characterizes the difference between parallel and perpendicular sizes of the cell. In this case the velocity takes the form:

$$\upsilon_x = -\sin x \cos\frac{y}{\alpha}; \tag{10}$$

$$v_y = \alpha \cos x \sin \frac{y}{\alpha}. \quad (11)$$

Note that at $y=0$ the $x$-velocity component is fixed and does not depend on aspect ratio $\alpha$.

We will consider the magnetic field $\vec{B}$ lying in the $xy$-plane. In this case, $\vec{B}$ can be expressed in terms of $z$-component of the vector potential $\vec{A} = A\vec{e}_z$ as follows:

$$B_x = \frac{\partial A}{\partial y}, \quad B_y = -\frac{\partial A}{\partial x}. \quad (12)$$

Note that curves $A = \text{const}$ coincide with the magnetic field lines.

We take the initial magnetic field $B_0$ constant an parallel to the $y$-axis, so that:

$$A = -B_0 x + a, \quad (13)$$

namely, $a|_{t=0} = 0$. In this case induction equation (1) in terms of potential $a$ will be written as follows:

$$\frac{\partial a}{\partial t} + v_x \frac{\partial a}{\partial x} + v_y \frac{\partial a}{\partial y} = B_0 v_x + \frac{1}{\text{Re}_m} \Delta a. \quad (14)$$

We solve this equation numerically for the periodic boundary condition relative to $x$: $a(\pi, y) = a(-\pi, y)$. As for dependence of $a$ with respect to $y$ we take also periodic boundary conditions continuing solution for positive $y$: $-\pi\alpha < y < +\pi\alpha$.

Equation (6) was solved numerically on the 2000 × 4000 grid $t_n = n\Delta t$, $x_i = -\pi + i\cdot\Delta x$, $y_j = -\pi + j\cdot\Delta y$; using a simple numerical scheme [30]:

$$\frac{a_{i,j}^{n+1} - a_{i,j}^n}{\Delta t} + (v_x)_{i,j}^n \frac{a_{i+1,j}^n - a_{i-1,j}^n}{2\Delta x} + (v_x)_{i,j}^n \frac{a_{i,j+1}^n - a_{i,j-1}^n}{2\Delta y} =$$
$$= B_0 (v_x)_{i,j}^n + \frac{1}{\text{Re}_m} \left( \frac{a_{i+1,j}^n - 2a_{i,j}^n + a_{i-1,j}^n}{\Delta x^2} + \frac{a_{i,j-1}^n - 2a_{i,j}^n + a_{i,j-1}^n}{\Delta y^2} \right)$$

Such schemes are stable for time steps $\Delta t = O^*(\text{Re}_m(\Delta x^2 + \Delta y^2))$ [33]. Usually we used $\Delta t = 2.5 \cdot 10^{-5}$.

Now present results of numerical simulation of equation (6) at $\text{Re}_m = \infty$. First of all we have varified that the magnetic field grows exponentially in time in the hyperbolic region which is defined in accordance with the Okubo-Weiss criterion [27, 28]:

$$\psi_{xx}\psi_{yy} - \psi_{xy}^2 < 0. \quad (15)$$

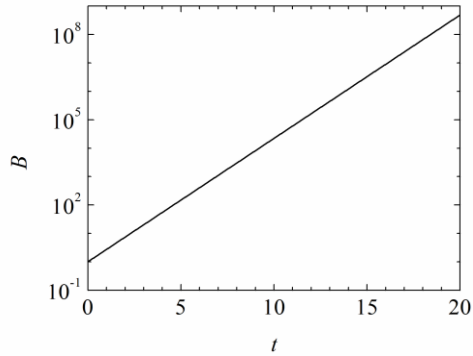

Figure 4. Evolution of the magnetic field for $\text{Re}_m = \infty$ at the maximal point $x=0$, $y=0$ (logarithmic scale).

We observed also that the magnetic field amplification in this region happens due to frozenness property of the magnetic field into plasma. In this case, the maximum of the growth rate is achieved in the corner $y=0$, $x=0$ which corresponds to the stationary hyperbolic point coinciding with the centers of the downward flow (see, Fig. 4). The magnetic field in this point is parallel to the downward convective flows. The vector potential A is shown on Fig. 6 at t=4.

We checked also that the maximal growth rate does not depend on aspect ratio, in complete correspondence with the general consideration presented above (see previous section, when $\text{Re}_m = \infty$).

At finite magnetic Reynolds number conductivity leads to destroy of the field frozenness that results in saturation of the exponential growth (see Fig.5).

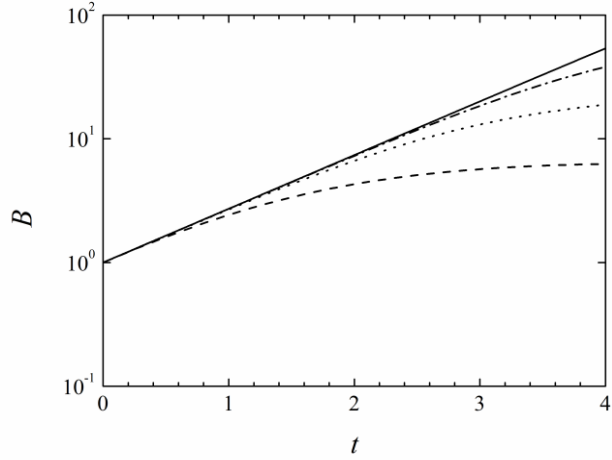

Figure 5. Evolution of the maximal magnetic field. Solid line shows $\text{Re}_m = \infty$, dashed line – $\text{Re}_m = 10$, dotted line – $\text{Re}_m = 10^2$, dot-dashed line – $\text{Re}_m = 10^3$.

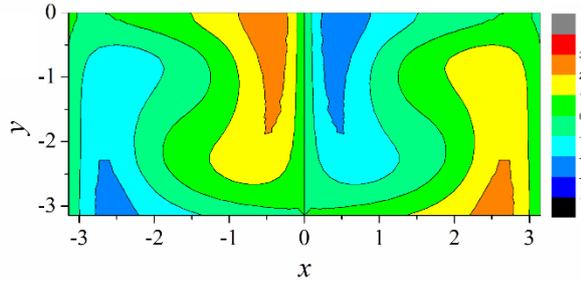

Figure 6. Vector potential of the magnetic field ($t = 4$).

At $\text{Re}_m \gg 1$ it is possible to estimate value of this field saturation in the magnetic filament. As was shown above the $B_x$ component near the maximal point is much smaller $B_y$ component that gives very different scales in horizontal and vertical directions. This means that in equation (3) for $B_y$ one can neglect by second derivative relative to $y$ in the Laplace operator and consider the velocity $v_x$ linear dependent on $x - x_2$:

$$\frac{\partial B_y}{\partial t} - \frac{\partial}{\partial x}(x B_y) = \text{Re}_m^{-1} \frac{\partial^2 B_y}{\partial x^2} \qquad 16)$$

where we designate $x - x_2$ as $x$ and $V = 1$. Saturation is reached on the stationary solution of this equation which after integration gives

$$B_y = B_{\max} \exp\left(-x^2 \text{Re}_m / 2\right) \qquad 17)$$

Value of $B_{\max}$ can be obtained using the magnetic field flux constancy. Approximately $B_{\max}$ is equal to [29]:

$$B_{\max} = 2.51 B_0 \text{Re}_m^{1/2}. \qquad (18)$$

According to [29] width $\delta$ of this stationary filament was estimated as $L \text{Re}_m^{-1/2}$.

## 5. Three-dimensional effects in magnetic filamentation

As we saw in the previous section for convection of the roll type appearance of magnetic filaments takes place in a small vicinity of the stationary hyperbolic point $y = 0$, $x = x_2 = 0$. In this region the velocity in dimensionless variables can be written as following: $v_x = -x$ and $v_y = y$. When magnetic viscosity is small enough and its influence is not too essential the $B_x$ component tends to zero exponentially in time but $B_y$ component grows exponetially. These filaments in this case represent a special kind of attractors. Account of finite magnetic viscosity leads to saturation of this grows and formation of stationary filaments parallel to the downward flow. At $\text{Re}_m \gg 1$, thus in the three-dimensional geometry these filaments form the whole plane of the small width. Consider their stability relative to $z$-coordinate fluctuations taking into account $B_z$ component. At $\text{Re}_m = \infty$ equation for $B_z$ has the form

$$\frac{\partial B_z}{\partial t} + (\vec{v} \cdot \vec{\nabla})B_z = 0; \tag{19}$$

where the velocity has two components with $v_z = 0$. Thus, the $B_z$ component represents passive scalar and can not change on the filaments. Moreover the viscous term, evidently, provides its dissipation. Therefore filaments in the three-dimensional case with planar downward flow has a tendency to be parallel to this surface.

However, the real situation, for instance, for solar cells has some peculiarities different from those for the roll-type convection. It is seen already from experimental data presented in Fig. 1 and Fig.2.

Firstly, solar filaments in the convective zone in their majority are concentrated near interfaces between convective cells and oriented correspondingly to downward flows [26]. Secondly, filaments are rare at the centers of convective cells. Inset of Fig.1, Fig. 2 for single cell shows in more details such features.

We will give explanation to these facts. First of all, remind that for the convective zone $\text{Re}_m$ is of order $10^6$ that allows one to use the frozenness equation:

$$\frac{\partial \vec{B}}{\partial t} = \vec{\nabla} \times (\vec{v} \times \vec{B}). \tag{20}$$

It is not difficult to get that the projection of this equation on the top convective surface ($z = 0$) gives for the normal ($\parallel \hat{z}$) component of $\vec{B}$ the following equation of motion:

$$\frac{\partial B_z}{\partial t} + (\vec{v}_\perp \cdot \vec{\nabla})B_z = -B_z \text{div}\,\vec{v}_\perp. \tag{21}$$

This equation is derived by means of incompressibility condition

$$\text{div}\,\vec{v}_\perp + \frac{\partial v_z}{\partial z} = 0 \tag{22}$$

and using boundary condition $v_z = 0$ at $z = 0$. The latter means that surface $z = 0$ is supposed not deformed, as it was considered in the 2D case. Therefore equation (21) represents an analogue of equation (4). Then by the same reason as before, at the flow source center $\vec{r} = \vec{r}_0$ the perpendicular velocity component $\vec{v}_\perp$ is equal to zero but its divergence there will be positive definite: $\text{div}\,\vec{v}_\perp > 0$. This provides exponential decrease in time of the normal component $B_z$ at the cell sourse center. However, at the interface with a neighboring cell (this is starting positions of a downward flow) we have the opposite situation: the transverse velocity vanishes but its divergence there takes negative values. This means that component $B_z$ will grow exponentially in time at the interface. The growth rate $\gamma$ equal to $|\text{div}\,\vec{v}_\perp|$ along the interface will be a function of its position. This means that some places at the interface will undergo bigger amplification of $B_z$ than in another regions of the interface. This inhomogeneity in the growth rate will cause filaments to form in regions of maximum enhancement while maintaining their orientation along the interface. As a result, the filaments will be quasi-flat as, for instance, it is seen in Fig.2. Note also that amplification does not depend on the sign of the normal component $B_z$: two kinds of filaments separated from each other are possible with different polarity of magnetic field (such situation can happen in the 3D case only).

Like in the 2D case, the second term in equation (19) describes advection of $B_z$ component towards the interface.

Thus, in the kinematic approximation we have two factors leading to the formation of filaments: advection and exponential growth of filaments at the interface. This explain experimental observations that solar filaments in the convective zone in their majority are concentrated near interfaces between convective cells and oriented correspondingly to downward flows. Magnetic filaments as it is seen from Fig.1 and Fig.2 are rare at cell centers because of advection of magnetic fields towards periphery due to the frozenness property.

As for magnetic field values in filaments in the 3D case they will have the same order of magnitude as in (18).

6. **Feedback**

Let us consider the Navier – Stokes equation in the presence of a magnetic field in the Boussinesq approximation

$$\frac{\partial \vec{\upsilon}}{\partial t} + (\vec{\upsilon} \cdot \vec{\nabla})\vec{\upsilon} = -\vec{\nabla}p + \nu R_a T\vec{n} + [[\vec{\nabla} \times \vec{B}] \times \vec{B}] + \nu \Delta \vec{\upsilon}, \quad (23)$$

where the magnetic field is written in Alfvenic units: $\vec{B}/\sqrt{4\pi\rho} \to \vec{B}$, i.e. has the dimension of speed, $R_a$ is the Rayleigh number, $T$ is the deviation of the temperature from its linear in $z$ dependence, $\vec{n}$ - unit vector along the vertical $z$, $\nu$ - kinematic velocity, $p$ - pressure.

In this case, we will assume temperature fluctuation vanishes at the boundary $z = 0$: $T|_{z=0} = 0$. Another boundary condition is written in the form of a continuity condition for the projection of the momentum flux density onto the normal:

$$\{\sigma_{ik} n_k\} = 0, \quad (24)$$

where curly brackets mean a jump at $z = 0$, and

$$\sigma_{ik} = -\upsilon_i \upsilon_k - (p + B^2/2)\delta_{ik} + B_i B_k + \nu\left(\frac{\partial \upsilon_i}{\partial x_k} + \frac{\partial \upsilon_k}{\partial x_i}\right). \quad (25)$$

Since the surface $z = 0$ is fixed, then the first term in (25) after substitution into (24) drops out due to $\upsilon_z|_{z=0} = 0$. It also follows that derivative $\vec{\nabla}_\perp \upsilon_z = 0$ (where $\vec{\nabla}_\perp$ is the gradient along the surface $z = 0$).

In the two-dimensional case, (24) implies two relations:

$$\{-B_x^2/2\} - p + 2\nu \frac{\partial \upsilon_y}{\partial y} = 0, \quad (26)$$

$$\{B_x\} B_y + \nu \frac{\partial \upsilon_x}{\partial y} = 0. \quad (27)$$

In these expressions, we took into account that the jump in the normal component of the magnetic field

$$\{B_y\} = 0. \quad (28)$$

As we saw earlier, in the region of magnetic field formation filament in case of high conductivity ($\text{Re}_m \gg 1$ [34]) component $|B_y| \gg |B_x|$ on boundary at $y = 0$. In this case, the relations (26, 27) at $\nu \neq 0$ approximately can be written as

$$-p + 2\nu \frac{\partial \upsilon_y}{\partial y} = 0, \quad (29)$$

$$\frac{\partial \upsilon_x}{\partial y} = 0. \quad (30)$$

In this case the equation (23) for $\upsilon_x$ at $y = 0$ takes the form

$$\frac{\partial \upsilon_x}{\partial t} + \upsilon_x \frac{\partial}{\partial x} \upsilon_x = -B_y \frac{\partial}{\partial x} B_y + 3\nu \frac{\partial^2}{\partial x^2} \upsilon_x + \nu \frac{\partial^2}{\partial y^2} \upsilon_x. \quad (31)$$

Far from filaments where $B_y \approx 0$ the convection flow should be independent on time, i.e. the last term in (31) only can provide such a stationary flow. In this case, the ratio between inertial term and viscous ne is of order of Reynolds number Re. According to [32, 34] it is of order of $10^2$. When we approach magnetic filaments this ratio is assumed valid. The compensation of the inertial term (second term in (31)) is due to gradient of magnetic pressure in filament. Thus the Hartmann number Ha defined as a ratio of electromagnetic force to the viscous force is of order of Reynolds number Re.

In the special case of zero viscosity, the relations (26, 27) are also significantly simplified:

$$\{B_x\} = 0,$$
$$p = 0 \text{ at } y = 0. \quad (32)$$

From (26, 27) it is also clear that for $\nu \to 0$ the longitudinal component of the magnetic field is also turns out to be small.

Thus, on the surface $y = 0$ with zero values of kinematic and magnetic viscosity and neglecting the transverse component magnetic field $B_x$ equation for the field velocity component $\upsilon_x$ and the components of the magnetic field $B_y$ turn out to be closed:

$$\frac{\partial v_x}{\partial t} + v_x \frac{\partial v_x}{\partial x} = -B_y \frac{\partial B_y}{\partial x}, \quad (33)$$

$$\frac{\partial B_y}{\partial t} + v_x \frac{\partial B_y}{\partial x} = -B_y \frac{\partial v_x}{\partial x}. \quad (34)$$

It is easy to see that the sum and difference $w_\pm = v_x \pm B_y$ obey two independent Hopf equations:

$$\frac{\partial w_\pm}{\partial t} + w_\pm \frac{\partial w_\pm}{\partial x} = 0. \quad (35)$$

In the case of a convective zone, at the initial moment we must assume that the average kinetic energy density at $t = 0$ significantly exceeds the magnetic energy density:

$$\frac{\langle v_x^2 \rangle}{2} \gg \frac{B_0^2}{2}. \quad (36)$$

According to the equation (33) one can see that growing magnetic field $B$ due to the magnetic pressure gradient (the r.h.s. of (33)) prevents penetration of the flow in the hyperbolic region with its center at $y = 0$, $x = 0$. By this reason, the hyperbolic point will be moved towards the counter flow that provides inverse influence of the growing magnetic field on the convective flow. Because the filament region is small in comparison with the convective cell such a shift should not influence significantly the convection itself. It is evident that this mechanism gives that magnetic pressure $B^2/2$ is comparable with the mean kinetic energy density $\langle v^2 \rangle/2$.

## 7. Conclusion

In this paper we have analyzed a filamentation of the magnetic field in convective cells in the Sun within the kinematic approximation. This process is associated with the frozenness of the magnetic field into a medium with high conductivity that leads to compression of magnetic field lines and forming magnetic filaments. Based on the general consideration of the convection top flows only, without knowledge of the cell structure, we demonstrate that the magnetic field intensifies in the regions of downward flows in both two-dimensional and three-dimensional convective cells. These regions of the hyperbolic type play a role of a specific attractor for the magnetic field. This theoretical analysis was confirmed by numerical simulations for 2D convective cells of the roll-type. Without dissipation, the magnetic field grows exponentially in time and attains its maximal value at the hyperbolic point where the growth rate does not depend on the aspect ratio between horizontal and vertical scales of the cell. This increase due to compression of the magnetic filaments is saturated due to the natural limitation associated with finite plasma conductivity when the maximum magnitude of the magnetic field is of the order of the root square of the magnetic Reynolds number. Another effect of saturation of the magnetic field values is connected with feedback of the growing field on the convective flows. These both effects for the Sun convective zone give for the maximal magnetic field values in filaments the same order of magnitude about 1 kG. Based on the stability analysis we have explained why downward flows influence magnetic filaments by making them more flat with orientation along interfaces between convective cells.

The authors thank V.V. Krasnoselskykh, D.D. Sokoloff and A.S. Shibalova for fruitful discussions and a number of valuable remarks.

Research was supported by Russian Science Foundation, grant number 19-72-30028.